\documentclass[aps,pre,twocolumn,superscriptaddress,showpacs]{revtex4}
\usepackage{graphicx}

\begin{document}
\title{Percolation of randomly distributed growing clusters: Finite Size Scaling and Critical Exponents}
\author{N. Tsakiris} \author{M. Maragakis} \author{K. Kosmidis} \author{P. Argyrakis}
\affiliation{Physics Department, Aristotle University of Thessaloniki,Thessaloniki 54124,Greece}

\date{\today}
\begin{abstract}
We study the percolation properties of the growing clusters model. In this model, a number of seeds placed on random locations on a lattice are allowed to grow with a constant velocity to form clusters. When two or more clusters eventually touch each other they immediately stop their growth. The model exhibits a discontinuous transition for very low values of the seed concentration $p$ and a second, non-trivial continuous phase transition for intermediate $p$ values. Here we study in detail this continuous transition that separates a phase of finite clusters from a phase characterized by the presence of a giant component. Using finite size scaling and large scale Monte Carlo simulations we determine the value of the percolation threshold where the giant component first appears, and the critical exponents that characterize the transition. We find that the transition belongs to a different universality class from the standard percolation transition. 
\end{abstract}
\pacs{64.60.ah, 61.43.Bn, 05.70.Fh, 81.05.Rm}
\maketitle

\section{Introduction}
Percolation represents a paradigmatic model of a geometric phase transition \cite{Bunde_BOOK,Stauffer,Newman_Ziff1,Newman_Ziff2,Isichenko,HoshKop,Kirkpatrik}. It has been widely studied and has numerous applications in many areas of physics \cite{Lorenz_Ziff,ziff2009explosive,lorenz1998universality,Wilkinson,Degennes,NewmanWatts,Sandler,Trugman,Cohen,Stauffer2}. Its importance and practical applications are described in detail elsewhere, see for example \cite{Stauffer}. Here, we will present for the sake of clarity and completeness, some necessary definitions of important quantities that commonly appear in the literature.
In the classical site percolation model, the sites of  a square lattice are randomly occupied with particles with probability $p$, or remain empty with probability $1-p$. Neighboring occupied sites are considered to belong to the same cluster and percolation theory simply deals with the number and properties of these clusters. When the occupation probability $p$ is small, the occupied sites are either isolated or form very small clusters. On the other hand, for large $p$ there are a lot of occupied sites that have formed one large cluster and it is possible to find several paths of occupied sites which a walker can use to move from one side of the lattice to the other. In this latter case, it is said that a giant component of connected sites exists in the lattice. This component does not appear in a gradual ``linear'' way with increasing $p$. It appears suddenly at a critical occupation probability $p_{c}$. Below $p_c$ there are only small clusters and even if we increase the lattice size considerably, these clusters remain small, i.e. the largest cluster does not depend on the system size. Above $p_c$, suddenly, small clusters join together to form a large cluster whose  size scales with system size. Hence, the term giant component or infinite cluster which is very common in the literature \cite{Bunde_BOOK}. 

In percolation, $p$ plays the same role as the temperature in thermal phase transitions, i.e. that of the control parameter, while the order parameter  is the probability $P_{\infty}$ that a site belongs to the infinite cluster. For $p>p_c$, $P_{\infty}$ increases with $p$ by a power law
\begin{equation}
P_{\infty} \sim (p-p_c)^{\beta}
\label{eq1}
\end{equation}  
Other important quantities are the correlation length $\xi$ which is defined as the mean distance between two sites on the same finite cluster and the mean number of sites $S$ of a finite cluster. When $p$ approaches $p_c$, $\xi$ increases as 
\begin{equation}
\xi \sim (p-p_c)^{-\nu}
\label{eq2}
\end{equation}
The mean number of sites $S$ of a finite cluster also diverges at $p_c$
\begin{equation}
S \sim (p-p_c)^{-\gamma}
\label{eq3}
\end{equation} 
The critical exponents $\beta,\nu$ and $\gamma$ describe the critical behavior associated with the percolation transition and are universal. They do not depend on the structure of the lattice (e.g., square or triangular) or on the type of percolation (site, bond or even continuum) \cite{Stauffer}.

In this paper we study numerically the percolation properties of the growing clusters model \cite{Tsakiris} which we describe in Sec. \ref{sec:model}. We focus on the intermediate concentration regime and find that the model exhibits a non-trivial percolation transition which belongs in a different universality class from standard percolation. We determine quite accurately the position of $p_c$, and the values of the critical exponents associated with this transition. 

\section{The Growing Clusters Model}
\label{sec:model}

\begin{figure}
\begin{center}
\includegraphics[width=8.5cm,height=8.5cm]{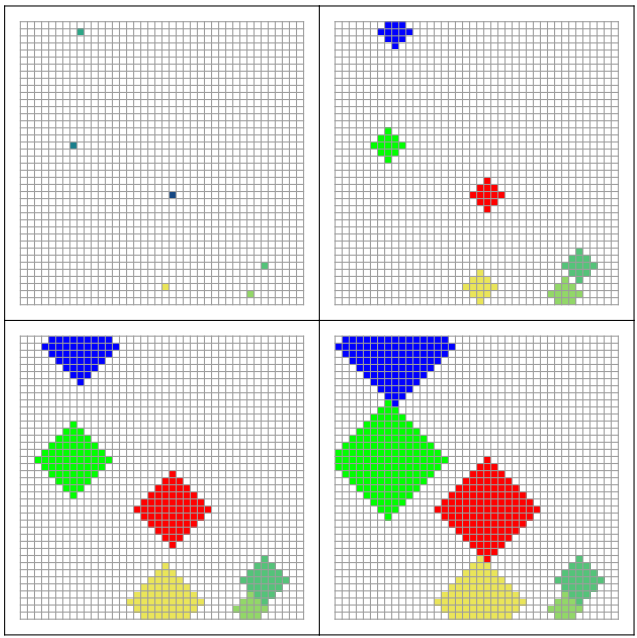}
\end{center}
\caption{Top Left: At time $t=0$ Monte Carlo Steps (MCS) 6 seeds are randomly placed on a $40 \times 40$ lattice. Top Right: After $t=3$ MCS there are 6 evolving clusters. Bottom Left: After $t=6$ MCS there are 4 evolving clusters, as 2 clusters have touched each other and stopped growing. Bottom Right: Final state of the system. There are no evolving clusters and 3 stable clusters have been formed.}
\label{graph1}
\end{figure}

The growing clusters model is presented in detail in \cite{Tsakiris}. Here, we provide a  brief description of it. A square lattice of size $L$ is randomly populated with ``seeds'' with probability $p$. These seeds are allowed to grow to clusters isotropically and stop when they touch another growing aggregate, see fig. \ref{graph1} for a schematic of the system evolution. At every time step, i.e. one Monte Carlo Step (MCS) all seeds are investigated once for the possibility to grow in size instantaneously in all neighboring sites. Investigation sequence is random in order. Each seed is allowed to grow its periphery by one layer (increase the radius by one) provided that there is no overlapping with other growing seeds. Thus, each seed becomes an evolving cluster. As soon as two, or more, clusters touch each other, the growth of all of the adjoined clusters stop. 
Neighboring sites are considered to belong in the same stable cluster. To study the model we perform large scale Monte Carlo simulations as described in \cite{Tsakiris} and we monitor the properties of the clusters that are formed.

\section{Finite Size Scaling}
Equations \ref{eq1}-\ref{eq3} are valid for infinite systems close to the critical threshold. In practice, however, it is not possible to use them to calculate the critical exponents with considerable accuracy due to finite size effects. Thus, one has to resort to finite size scaling \cite{Stauffer} techniques. Due to the finite size of the lattices that can be simulated the order parameter is expected to depend on the system size. Assuming that we are close to the critical threshold so that the correlation length $\xi$ is comparable to the system size $L$ it can be shown that the probability $P_{max}$ that a site belongs to the largest cluster follows the scaling relation:
\begin{equation}
P_{max}= L^{-\beta/\nu} F[L^{1/\nu} (p-p_c)]
\label{eq4}
\end{equation}
where we have deliberately used the notation $P_{max}$ instead of $P_{\infty}$ in order to emphasize the finiteness of the systems. Here $F$ is  a suitable scaling function. Similarly, any other quantity varying as $\mid p-p_c\mid ^x$ is expected to scale similar to eq. \ref{eq4} with $\beta$ replaced by the appropriate exponent $x$ and with a different scaling function $F$.
At $p=p_c$, these quantities are expected to scale as power law since the scaling function reduces to a simple proportionality constant. This result gives a way to determine the critical exponents.
One important characteristic of the standard percolation transition is that exactly at the critical point, the largest cluster has a fractal geometry meaning that the mass of the largest cluster $S_1$ scales with the system size as $S_1 \sim L^{d_f}$, where the fractal dimension $d_f$ is known to be equal to 91/48. Since $S_1=P_{max} L^d$, where $d$ is the dimensionality of the space($d=2$ in the lattice case)  one can easily derive a scaling law relating $d_f,\beta$ and $\nu$, namely $d_f=d-\beta/\nu$.  
 
\section {Results and Discussion}
To obtain an indication of the system critical behavior, we start by monitoring the size of the largest cluster $S_{1}$ as a function of the initial ``seed'' probability $p$, see \cite{Tsakiris} and fig. 2 therein. There, it is evident that the system exhibits two phase transitions: One very sharp transition at $p=0$ and a second, smoother, transition around $p \simeq 0.5$. The first transition is discussed in detail at \cite{Tsakiris} and characterized by the fact that the size of the largest cluster $S_1$, which is the order parameter of the system, has a discontinuity for $p=0$ and, thus, the system exhibits a sharp, albeit artificial, first order phase transition. 

The second phase transition turns out to be rather interesting. It is reminiscent of the classical percolation transition and it is important to clarify the extent of this similarity. Thus, we simulate systems of several different sizes and for several different initial seed concentrations. After allowing the growth process to complete and the systems to reach their final states, we study the probability $P_{max}$ that a randomly chosen site belongs to the largest cluster. When there are only few initial seeds, after the evolution of the system is completed, the clusters formed  are small and isolated. However, we expect that with increasing $p$ when a large number of seeds is introduced, the growth process will lead to the formation of a large spanning cluster. This is the classic behavior seen in a system which undergoes a continuous phase transition. In such cases, we can use finite size scaling to determine the position of the critical concentration, $p_c$, and the critical exponents.
\begin{figure}
\begin{center}
\includegraphics[width=8.5cm,height=8.5cm]{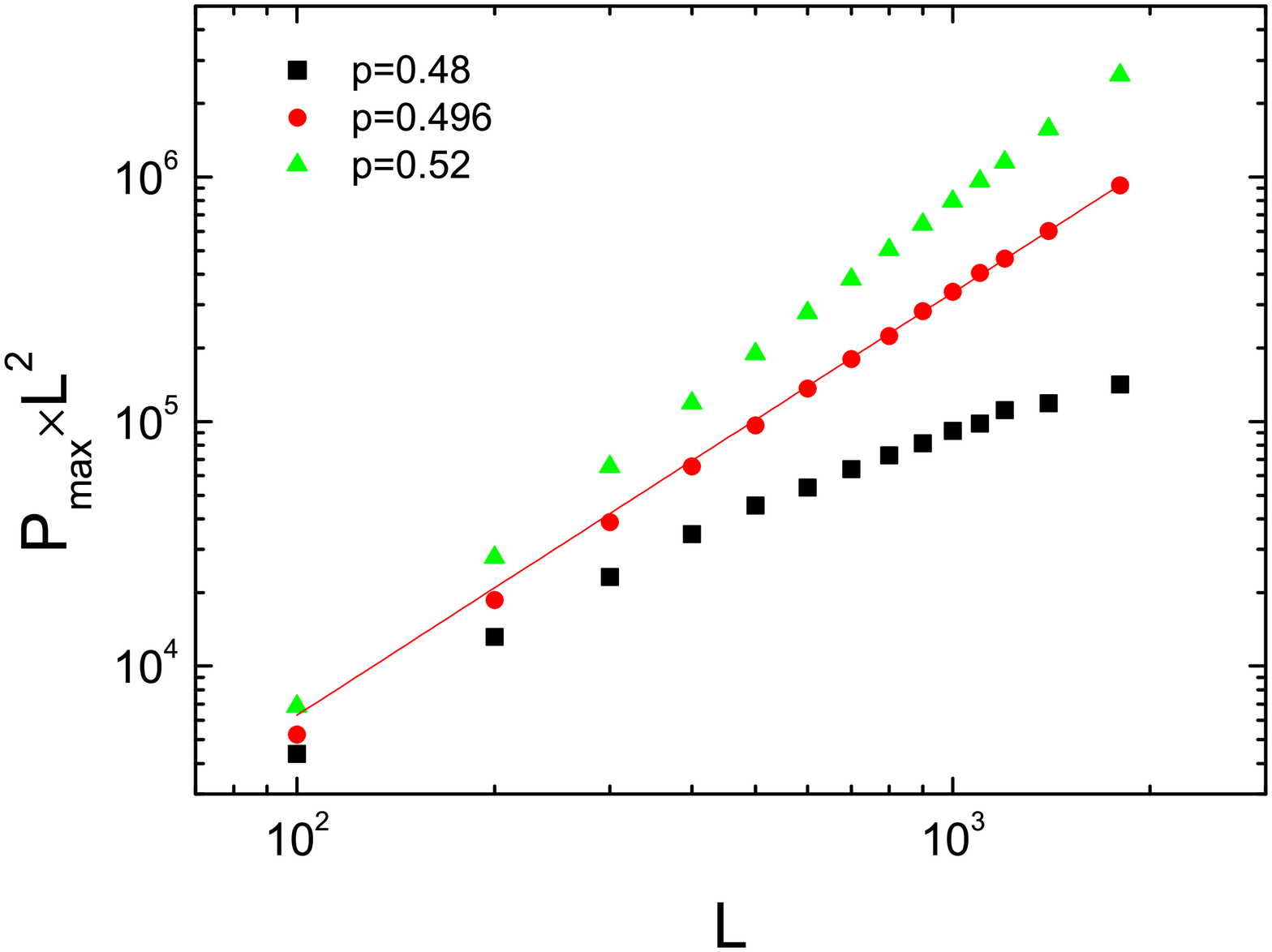}
\end{center}
\caption{Size of the largest cluster, $S_1= P_{max}L^2$, as a function of the lattice size $L$ for three different initial concentrations, $p=0.48, 0.496$ and $0.52$(black squares, red dots and green triangles respectively). Points are Monte Carlo Simulation results and the straight line is a power law fit with slope 1.79.}
\label{fig1}
\end{figure}
Initially, we are interested in the geometry of the largest cluster at criticality and its fractal dimension $d_f$. In fig. \ref{fig1} we plot the size of the largest cluster $S_1= P_{max}L^2$ as a function of the lattice size $L$ for three different initial concentrations, namely $p=0.48, 0.496$ and $0.52$ (black squares, red dots and green triangles respectively). For $p=0.48$ (squares) we observe a downward bending curve which clearly indicates that $p=0.48$ is below $p_c$ as $S_1$ does not scale with the system size for large $L$. For $p=0.496$ (circles) we observe a straight line implying a power law scaling as expected for $p=p_c$. This value for the critical point agrees reasonably well with our preliminary estimation for the critical point from the position of the maximum of the second largest cluster \cite{Tsakiris}. For $p=0.52$ (triangles) we observe a curve which is bending slightly upwards for large $L$, indicating that we are above $p_c$, although admittedly for the sizes of lattices simulated here is quite difficult to observe the bending for $p=0.52$. We will, however, use another method below to confirm that $p_c=0.496$ for the growing clusters model. We calculate the fractal dimension, $d_f$, from the slope of the straight line at $p_c$ and find $d_f \simeq 1.79$.  

\begin{figure}
\begin{center}
\includegraphics[width=8.5cm,height=8.5cm]{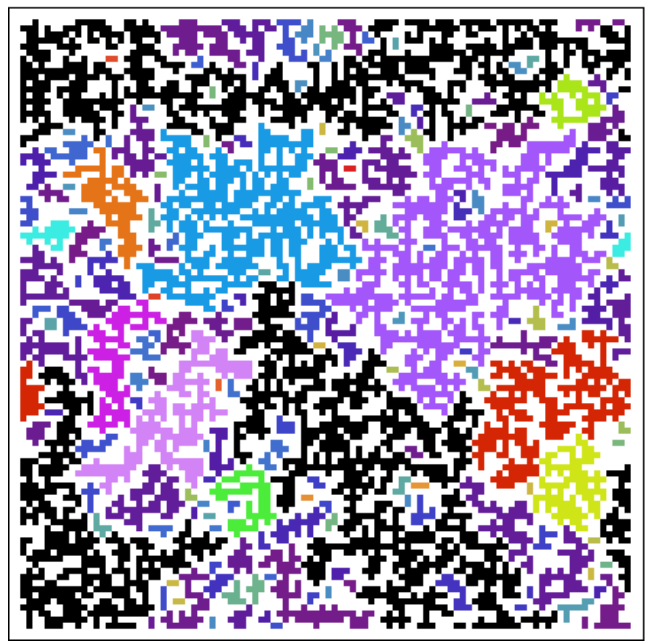}
\end{center}
\caption{Snapshot of the final state of a system with $L=100$ with initial seed concentration $p =0.496$ which is equal to the critical concentration ($p=p_c$). Different colors correspond to different clusters. The final coverage is $p_f=0.538$. The largest cluster is shown in black color.}
\label{graph2}
\end{figure}  
Figure \ref{graph2} shows a snapshot of the clusters that have been formed in a random realization of a growing clusters system with $L=100$ at $p_c=0.496$ once the system has reached its steady state. Periodic boundary conditions have been used for the cluster labeling. The largest cluster is shown in black color.  

\begin{figure}
\begin{center}
\includegraphics[width=8.5cm,height=8.5cm]{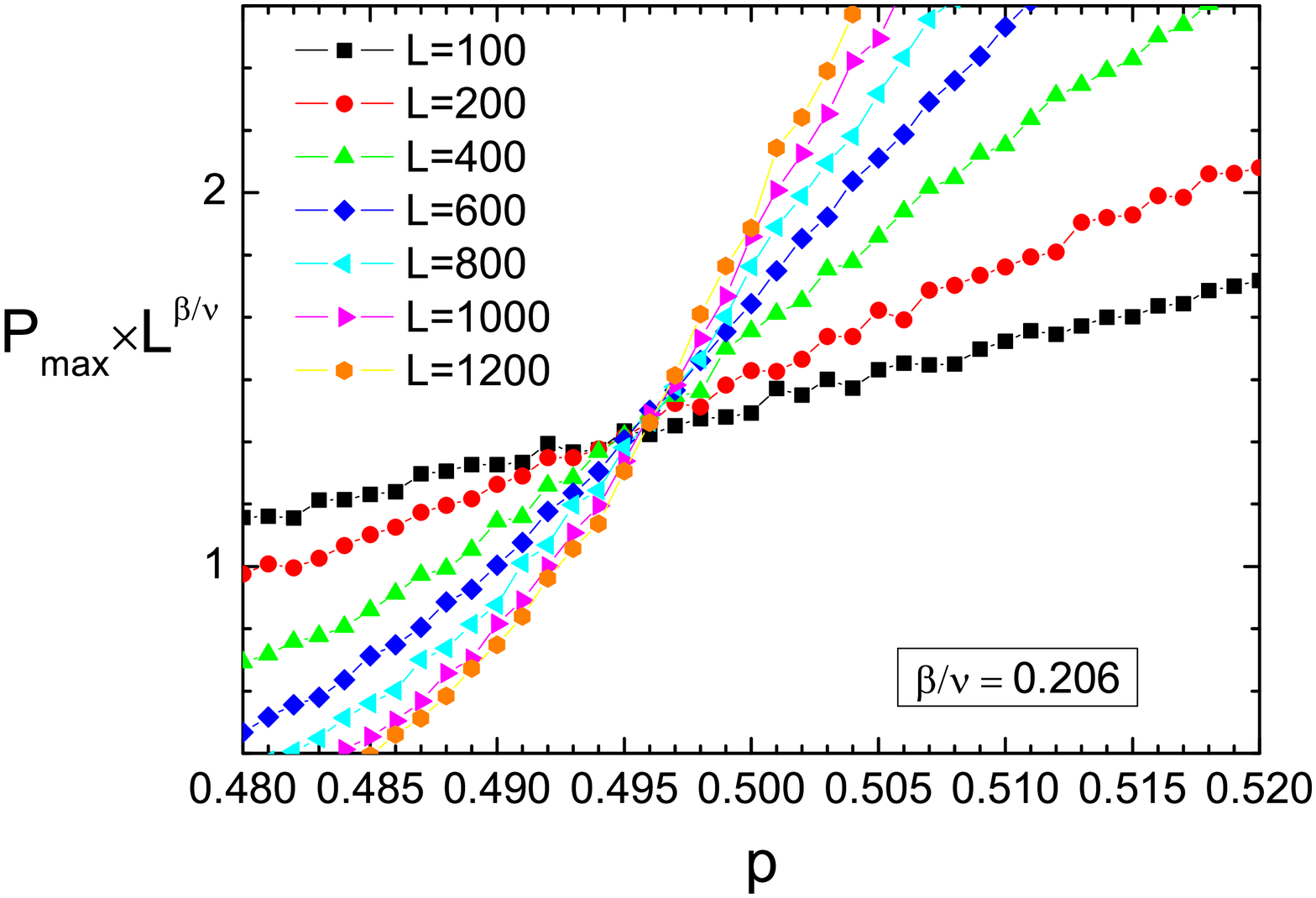}
\end{center}
\caption{$P_{max} L^{0.206}$ vs $p$ for seven different system sizes, namely $L=100,200,400,600,800,1000,1200$. The curves cross at $p_c=0.496$ in agreement with the scaling relation eq. \ref{eq4}.}
\label{fig2}
\end{figure} 
In order to determine more accurately the percolation threshold and the critical exponent $\beta/\nu$ ratio in the same time, we use eq. \ref{eq4}. We plot in fig. \ref{fig2} $P_{max} L^{\beta/\nu}$ vs $p$ for seven different system sizes, namely $L=100,200,400,600,800,1000,1200$, and we vary $\beta/\nu$ until all curves 
cross at one single point only. This is done for $\beta/\nu = 0.206$ and for $p=p_c \simeq 0.496$. This result is in excellent agreement with our previous estimation for the $d_f$ and the scaling relation $d_f=d-\beta/\nu$. It also allows  to determine the exact location of the critical point with more accuracy than the previous method.  
 
\begin{figure}
\begin{center}
\includegraphics[width=8.5cm,height=8.5cm]{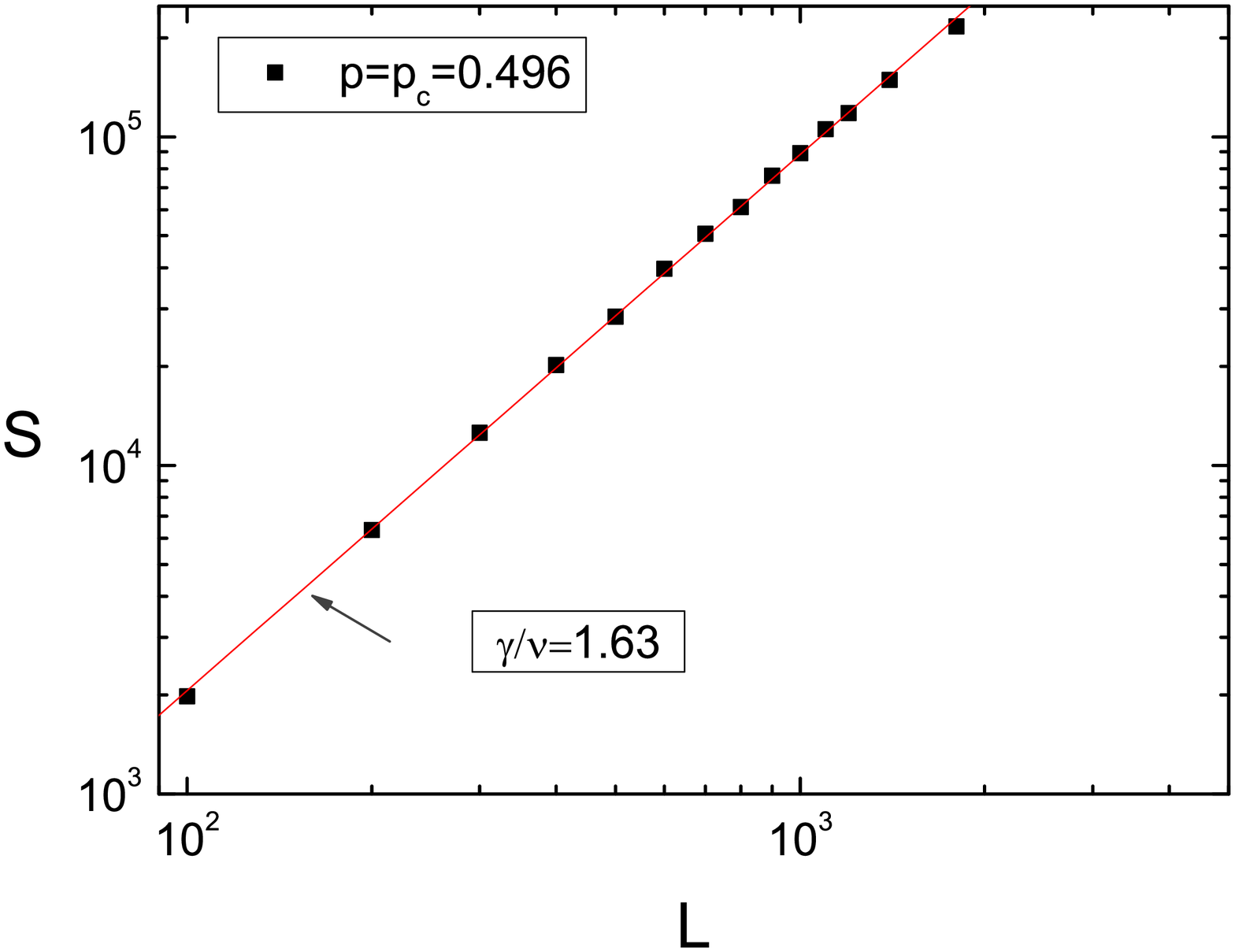}
\end{center}
\caption{Mean mass $S$ of the finite clusters as a function of $L$ for $p=p_c$. Points are Monte Carlo Simulation results. The straight line is a best fit to the simulation data and has a slope of 1.63}
\label{fig3}
\end{figure} 
Fig. \ref{fig3} shows a plot of the mean mass $S$ of all the finite clusters as a function of $L$ for $p=p_c$. At criticality, this quantity is expected to scale as $S \sim L^{\gamma/\nu} $ (see section III). From the slope of the straight line we estimate $\gamma/\nu =1.63$. This, as well as the previous result, are in very good agreement with the scaling relation $d\nu=\gamma +2\beta$. 

\begin{figure}
\begin{center}
\includegraphics[width=8.5cm,height=8.5cm]{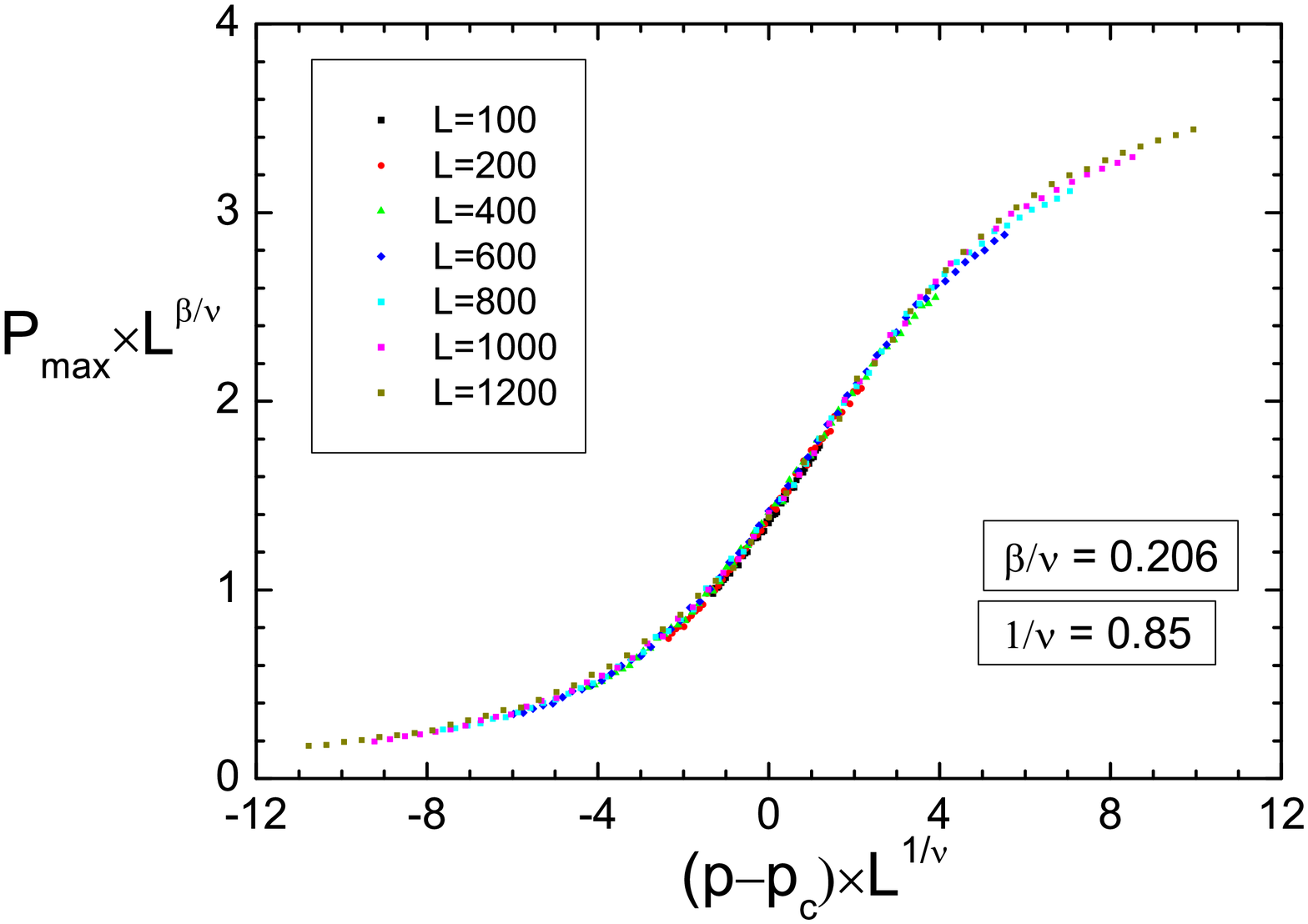}
\end{center}
\caption{$P_{max} L^{0.206}$ vs $(p-p_c) L^{0.85}$ for seven different system sizes $L=100,200,400,600,800,1000,1200$. Points are simulation data. The data collapse to one single curve. Thus we estimate $1/\nu=0.85$.}
\label{fig4}
\end{figure} 
Finally, we calculate the critical exponent $\nu$. In fig. \ref{fig4} we plot $P_{max} L^{0.206}$ vs $(p-p_c) L^{1/\nu}$ for seven different system sizes $L=100,200,400,600,800,1000,1200$, and we vary the exponent $1/\nu$ until all data collapse on one single curve. The data collapse, in agreement with eq.\ref{eq4} enables us to determine the critical exponent $\nu$ with considerable accuracy. We find that $1/\nu=0.85$.

We can determine other critical exponents from the scaling relations that are known to connect them. Below, for completeness, we present a table with the critical exponents of the growing clusters model in comparison to those of classical percolation. The difference in the critical exponents shows that the two models belong in a different universality class. Moreover, our calculated value of $\nu$ implies that the growing clusters model is also in a different universality class from the ``invasion percolation'' \cite{Wilkinson} model.

\begin{table}[h]
\begin{center}
\begin{tabular}{|c|c|c|}
\hline
Exponent & Class. Percolation & Growing Clusters \\
\hline
$\beta$ & 0.138 & 0.24  \\
\hline
$\gamma$ & 2.38 & 1.91 \\
\hline
$\nu$ & 1.33 & 1.17 \\
\hline
$\sigma=1/(\beta +\gamma)$ & 0.395 & 0.46 \\
\hline
$\tau=1+\sigma \nu d$ & 2.05 & 2.08 \\
\hline
$d_f$ & 1.89 & 1.79 \\
\hline
\end{tabular}
\caption{\label{tab1}Comparison of the critical exponents between the classical percolation transition and the transition of the growing clusters model. The exponents $\tau$ and $\sigma$ are associated with the cluster size distribution \cite{Bunde_BOOK}}.
\end{center}
\vspace{-0.6cm}
\end{table}

\section{Conclusions}
We have studied the growing clusters model and found that it exhibits two phase transitions with increasing seed concentration. A first order transition at $p=0$ and a continuous transition at $p_c=0.496$, separating a phase of isolated clusters for $p<p_c$ from a phase where a giant component is present for $p>p_c$. Using finite size scaling we have calculated the position of the phase transition and the critical exponents with considerable accuracy to establish that this transition belongs to a different universality class from the standard percolation transition.

{\it Acknowledgments}: This work was supported by the FP6 Project INTERCONY NMP4-CT-2006-033200.

\end{document}